\documentclass[runningheads]{llncs}

\usepackage{graphicx}
\usepackage[utf8]{inputenc}
\usepackage{multirow}
\usepackage{float}

\begin{document}

\title{Automatic Rodent Brain MRI Lesion Segmentation with Fully Convolutional Networks
}

\titlerunning{Automatic Rodent Brain MRI Lesion Segmentation with FCNs.}

\author{Juan Miguel Valverde\inst{1} \and
Artem Shatillo\inst{2} \and
Riccardo de Feo\inst{3} \and
Olli Gröhn\inst{1} \and
Alejandra Sierra\inst{1} \and
Jussi Tohka\inst{1}}

\authorrunning{J. M. Valverde et al.}

\institute{AI Virtanen Institute for Molecular Sciences, University of Eastern Finland, Kuopio, Finland \\
\email{\{juanmiguel.valverde,alejandra.sierralopez,olli.grohn,jussi.tohka\}@uef.fi}\\
\and
Charles River Discovery Services, Kuopio, Finland\\
\email{artem.shatillo@crl.com}
\and
Centro Fermi - Museo Storico della Fisica e Centro Studi e Ricerche Enrico Fermi, Rome, Italy\\
\email{riccardo.defeo@uniroma1.it}}

\maketitle

\begin{abstract}
Manual segmentation of rodent brain lesions from magnetic resonance images (MRIs) is an arduous, time-consuming and subjective task that is highly important in pre-clinical research. Several automatic methods have been developed for different human brain MRI segmentation, but little research has targeted automatic rodent lesion segmentation. The existing tools for performing automatic lesion segmentation in rodents are constrained by strict assumptions about the data. Deep learning has been successfully used for medical image segmentation. However, there has not been any deep learning approach specifically designed for tackling rodent brain lesion segmentation. In this work, we propose a novel Fully Convolutional Network (FCN), RatLesNet, for the aforementioned task. Our dataset consists of 131 T2-weighted rat brain scans from 4 different studies in which ischemic stroke was induced by transient middle cerebral artery occlusion. We compare our method with two other 3D FCNs originally developed for anatomical segmentation (VoxResNet and 3D-U-Net) with 5-fold cross-validation on a single study and a generalization test, where the training was done on a single study and testing on three remaining studies. 
The labels generated by our method were quantitatively and qualitatively better than the predictions of the compared methods. The average Dice coefficient achieved in the 5-fold cross-validation experiment with the proposed approach was 0.88, between 3.7\% and 38\% higher than the compared architectures. The presented architecture also outperformed the other FCNs at generalizing on different studies, achieving the average Dice coefficient of 0.79. 

\end{abstract}

\section{Introduction}
Medical image segmentation constitutes a bottleneck in research due to the time required to delineate regions of interest (ROI). In particular, pre-clinical studies involving animals can potentially produce 3D images in the order of hundreds, i.e., several animals with varied characteristics such as age, gender and strain at different time-points may be needed for a certain study.

Rodent brain lesion segmentation is a challenging problem subject to inter- and intra-operator variability. There are multiple factors that influence segmentation quality such as the characteristics of the lesion, the scan parameters and the annotators' knowledge and experience. Despite that, manual segmentation remains the de facto gold standard, decreasing the reproducibility of research.

Convolutional neural networks (CNNs) and, in particular, U-Net-like architectures \cite{unet} have recently become a popular choice for anatomical segmentation. There are numerous methods specifically designed for human brains to parcellate anatomical structures \cite{voxresnet,quicknat}, tumors \cite{braintumor} and lesions \cite{bratspaper}.  On contrary, animal brain segmentation algorithms based on CNNs are still infrequent \cite{shortreview}. As one example, Roy et al. showed that the same CNN can be used for extraction of human and mouse brains (i.e., skull-stripping) \cite{micebrain}.

Recently, few automatic approaches were developed to tackle rodent brain lesion segmentation, including statistical models \cite{stat}, thresholding \cite{thresholding} and level-sets \cite{levelsets}. However, these methods rely on strict assumptions such as the distribution of the data, or they are limited to use a single image modality. To the best of the authors' knowledge, CNNs have not yet been proposed for segmenting rat brain lesions.

In this work, we present RatLesNet, a 3D Fully Convolutional Network (FCN) to automatically segment rat brain lesions. Our dataset consists of a total of 131 Magnetic Resonance (MR) T2-weighted rat brain scans from 4 different studies with their corresponding lesion segmentation. Brain scans were acquired at different time-points after the lesions were caused by ischemic stroke induced by an occlusion of transient middle cerebral artery. Unlike other studies in which the segmentation algorithms rely on additional steps \cite{stat,thresholding,levelsets}, RatLesNet uses unprocessed MR-images that are not skull-stripped nor corrected for bias-field inhomogeneity. As well, we do not post-process the final predictions by filling holes or discarding small isolated clusters of voxels.

We compare RatLesNet with two other 3D FCNs performing internal five-fold cross-validation in one study of the dataset, and by training on volumes from that study to assess their generalization capability on the remaining 3 studies.

\begin{figure}
\includegraphics[width=\textwidth]{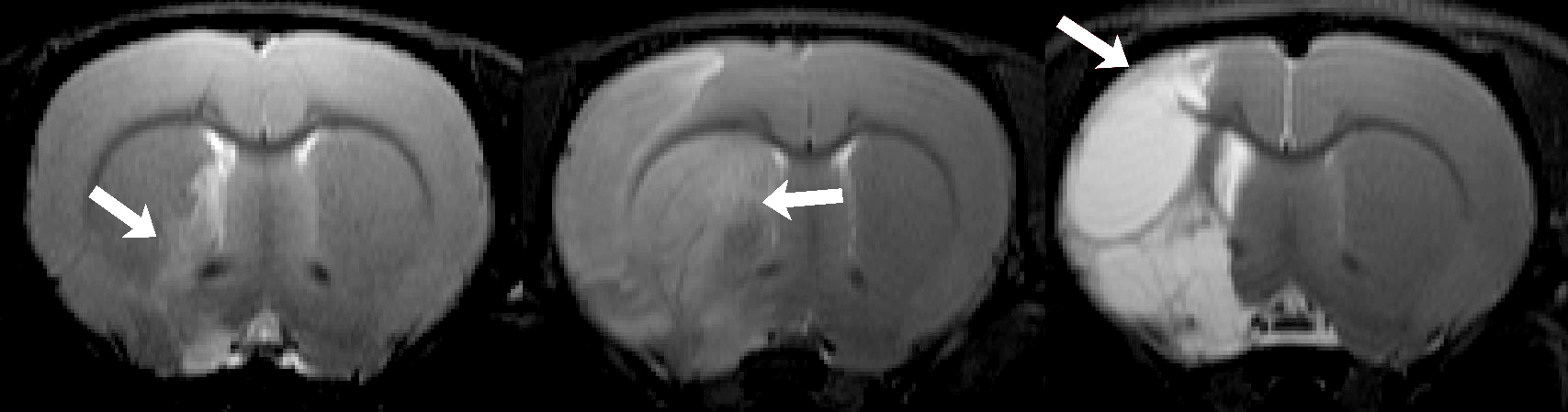}
\caption{Representative lesion progression of a rat at 2h, 24h and 35 days after MCA occlusion. The arrows point to the lesion, which appear hyperintense in T2-weighted images on the right hemisphere.} \label{fig1}
\end{figure}

\section{Materials and methods}

\textbf{Material}: The dataset consists of 131 MR brain scans of different adult male Wistar rats weighting between 250-300 gr. The data, provided by Charles River Laboratories\footnote{https://www.criver.com/}, are derived from four different studies: 03AUG2015 (21 scans), 03MAY2016 (45 scans), 02NOV2016 (48 scans) and 02OCT2017 (17 scans). Transient (120 min) focal cerebral ischemia was produced by middle cerebral artery occlusion (tMCAO) in the right hemisphere of the brain \cite{koizumimodel}. MR data acquisitions were performed at different time-points in a horizontal 7T magnet, more specifically at 2 hours (12 animals from 02NOV2016 study), 24 hours (12 animals from 02NOV2016 and all animals from 03MAY2016 and 02OCT2017) and 35 days (all animals from 03AUG2015 except 1) after the occlusion. 02NOV2016 and 03AUG2015 studies included 24 and 1 sham animals, respectively, that underwent through identical procedures, including anesthesia regime, but without the actual tMCAO occlusion. All animal experiments are carried out according to the National Institute of Health (NIH) guidelines for the care and use of laboratory animals, and approved by the National Animal Experiment Board, Finland. Multi-slice multi-echo sequence was used with the following parameters; TR = 2.5 s, 12 echo times (10-120 ms in 10 ms steps) and 4 averages. Eighteen (18) coronal slices of thickness 1 mm were acquired using a field-of-view of 30x30 mm$^2$ producing 256x256 imaging matrices. 

The T2-weighted MR images and their corresponding lesion segmentation were provided in form of NIfTI files. We chose a single study (02NOV2016) and performed an independent lesion segmentation to approximate inter-rate variability. The Dice coefficient \cite{dice1945measures} between our independent manual segmentation and the one provided was 0.73.

\begin{figure}
\includegraphics[width=\textwidth]{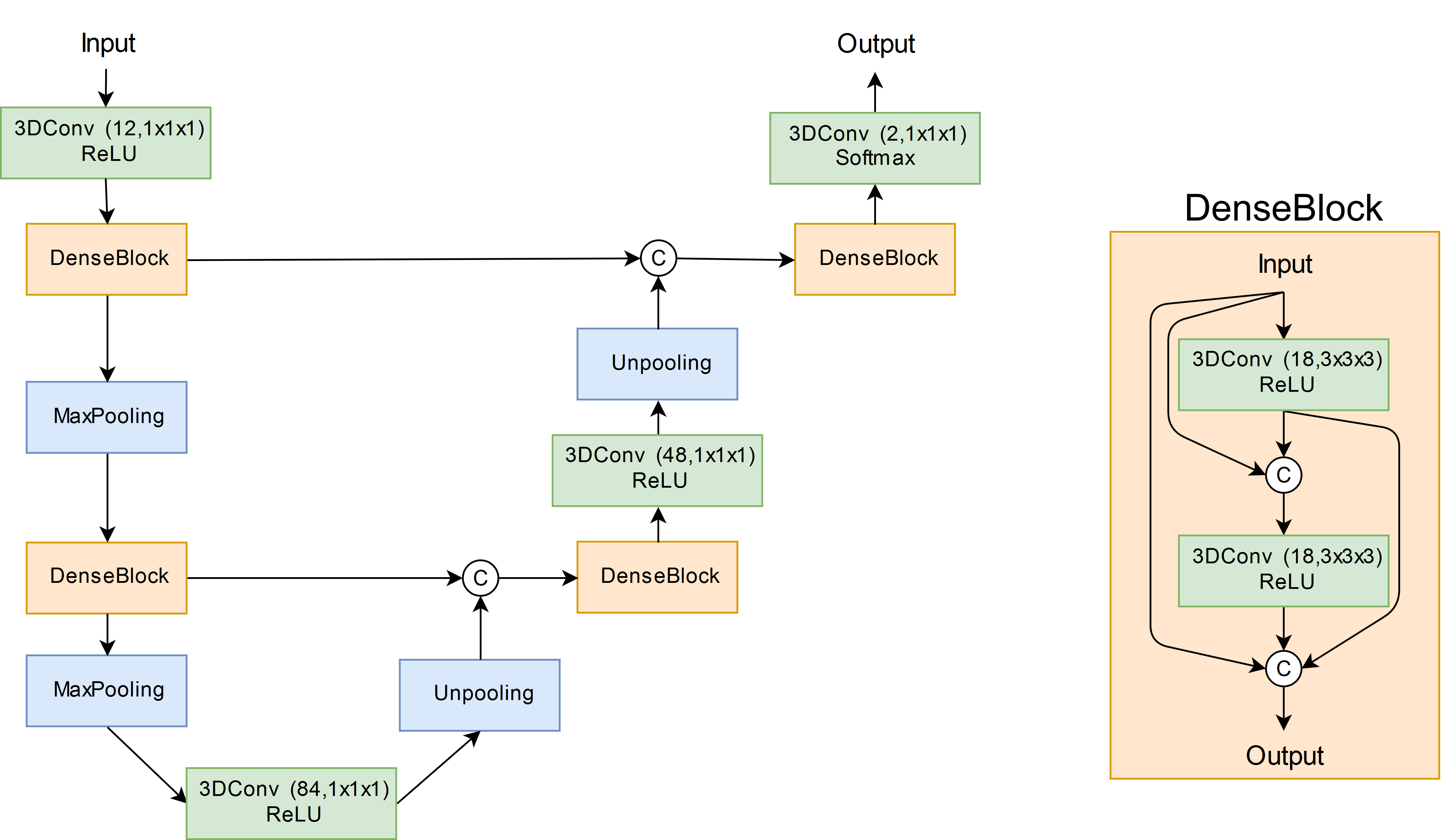}
\caption{Overview of the network architecture.} \label{network}
\end{figure}

\noindent \textbf{Network architecture}: RatLesNet's architecture (Figure \ref{network}) is composed by three types of blocks: 1) Dense blocks, 2) 3D convolutions with filter size of 1 followed by an activation function and 3) Max-pooling/Unpooling operations. Dense blocks encapsulate two 3D convolutions with filter size of 3 and ReLU activation functions concatenated with the outputs of the previous layers within the same block in a ResNet \cite{resnet} fashion. The number of channels increasing within the block (i.e. growth rate) is 18 and, unlike the original Dense blocks \cite{densenet}, these do not include a transition layer at the end. Due to the consequent widening of layers along the network, the external 3D convolutional layers with filter size of 1 are of special importance for adjusting the number of channels, similarly to traditional bottleneck layers. Max-pooling and unpooling layers are used to reduce and recover the spatial dimensions of the volumes, respectively. Max-pooling has a window size and stride of 2, and unpooling reuses the indices of the values from the max-pooling operation to place the new values back to their original location \cite{unpooling}. Finally, the input and output volumes of the network have identical spatial dimensions, and their number of channels correspond to the number of medical imaging modalities (here 1 that is T2) and the number of classes (here 2: lesion and non-lesion), respectively.

\noindent \textbf{Training}: The network was trained on entire T2-weighted images with a resolution of 256x256x18 voxels and a mini-batch size of 1. Image intensities in each volume were standardized by subtracting their mean and dividing them by their standard deviation. Cross-entropy loss function was minimized using Adam \cite{adam} with a starting learning rate of $10^{-5}$. Training lasted for a maximum of 1000 epochs or until the validation loss increased 5 consecutive times.

\section{Experiments and Results}
\textbf{Cross-validation}. Five-fold cross-validation was performed on the scans of the study that was segmented independently (02NOV2016) to assess the performance of the proposed RatLesNet method. Additionally, RatLesNet was compared to 3D U-Net \cite{3dunet} and VoxResNet \cite{voxresnet}, two 3D fully convolutional networks originally designed for anatomical segmentation. Both architectures were implemented according to the original design resulting in 19M parameters (3D U-Net) and 1.5M parameters (VoxResNet). In contrast,  our RatLesNet implementation has only 0.37M parameters, reducing the possibility of overfitting. The models that provided the reported results were trained with the best performing learning rate found.

\begin{table}[h]
\caption{Average 5-fold cross-validation Dice scores and standard deviation from 02NOV2016 dataset with and without sham animals (no brain lesion). The column Inter-operator refers to differences between the two independent manual segmentations by different operators.}\label{tab1}
\centering

\begin{tabular}{|l|c|c|c|c|c|}
\hline
\hspace{0.01in} Time-point \hspace{0.01in} & \hspace{0.01in} Shams \hspace{0.01in} & \hspace{0.01in} 3D U-Net \hspace{0.01in} & \hspace{0.01in} VoxResNet \hspace{0.01in} & \hspace{0.01in} RatLesNet \hspace{0.01in} & \hspace{0.01in} Inter-operator \hspace{0.01in}\\

\hline
    
\multirow{2}{*}{\hspace{0.01in} 2h} & Yes & \hspace{0.01in} $0.57 \pm 0.42$ \hspace{0.01in} & \hspace{0.01in} $0.80 \pm 0.23$ \hspace{0.01in} & \hspace{0.01in} $0.84 \pm 0.19$ \hspace{0.01in} & \hspace{0.01in} $0.84 \pm 0.19$ \hspace{0.01in} \\
                     & No & \hspace{0.01in} $0.17 \pm 0.13$ \hspace{0.01in} & \hspace{0.01in} $0.60 \pm 0.16$ \hspace{0.01in} & \hspace{0.01in} $0.67 \pm 0.12$ \hspace{0.01in} & \hspace{0.01in} $0.67 \pm 0.12$ \hspace{0.01in}  \\
\multirow{2}{*}{\hspace{0.01in} 24h} & Yes & \hspace{0.01in} $0.7 \pm 0.35$ \hspace{0.01in} & \hspace{0.01in} $0.89 \pm 0.18$ \hspace{0.01in} & \hspace{0.01in} $0.92 \pm 0.11$ \hspace{0.01in} & \hspace{0.01in} $0.90 \pm 0.12$ \hspace{0.01in} \\
                     & No & \hspace{0.01in} $0.43 \pm 0.13$ \hspace{0.01in} & \hspace{0.01in} $0.79 \pm 0.2$ \hspace{0.01in} & \hspace{0.01in} $0.85 \pm 0.11$ \hspace{0.01in} & \hspace{0.01in} $0.79 \pm 0.08$ \hspace{0.01in}  \\ \hline
\multirow{2}{*}{\hspace{0.01in} Average} & Yes & \hspace{0.01in} $0.64 \pm 0.39$ \hspace{0.01in} & \hspace{0.01in} $0.85 \pm 0.21$ \hspace{0.01in} & \hspace{0.01in} $0.88 \pm 0.16$ \hspace{0.01in} & \hspace{0.01in} $0.87 \pm 0.16$ \hspace{0.01in} \\
                     & No & \hspace{0.01in} $0.30 \pm 0.26$ \hspace{0.01in} & \hspace{0.01in} $0.70 \pm 0.20$ \hspace{0.01in} & \hspace{0.01in} $0.76 \pm 0.14$ \hspace{0.01in} & \hspace{0.01in} $0.73 \pm 0.12$ \hspace{0.01in} \\
\hline
\end{tabular}
\label{tab:table1}
\end{table}

Our RatLesNet provided better quantitative and qualitative results than 3D U-Net \cite{3dunet} and VoxResNet \cite{voxresnet}. Table \ref{tab:table1} compares the mean Dice coefficients and their variability across the implemented networks. As brain lesion's appearance varies depending on the time passed since the lesion was caused (see Figure \ref{fig1}), Dice coefficients were averaged separately.  Furthermore, due to the incorporation of sham animals (rats without lesion), two averages are provided: one that includes all animals and one that excludes rats without lesion. The methods recognized well the brains with no lesions and therefore the average Dice coefficients and their standard deviations were higher when shams were included. Due to the pronounced class imbalance, the network tends to classify voxels as non-lesion. On average, the number of lesion voxels in our independent segmentation in 2h and 24h time-points were 6060 ($0.51\%$ of all voxels) and 21333 ($1.81\%$ of all voxels), respectively. Additionally, 2h scans are more troublesome to segment because the lesion is considerably smaller and some affected areas resemble healthy tissue.

\begin{figure}
\centering
\includegraphics[width=0.6\textwidth]{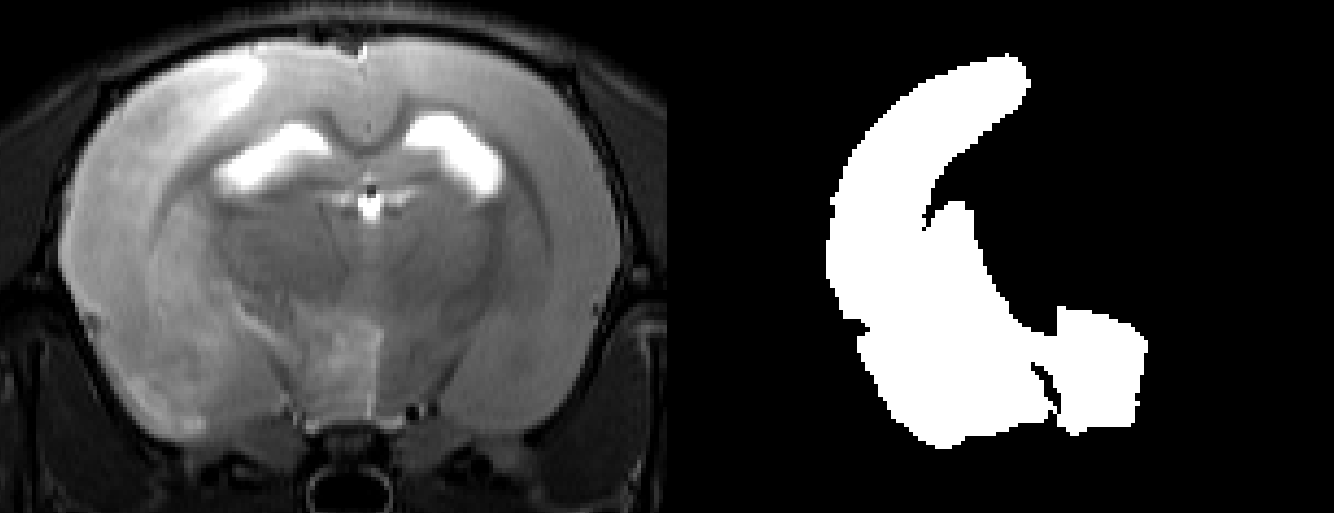}
\includegraphics[width=0.9\textwidth]{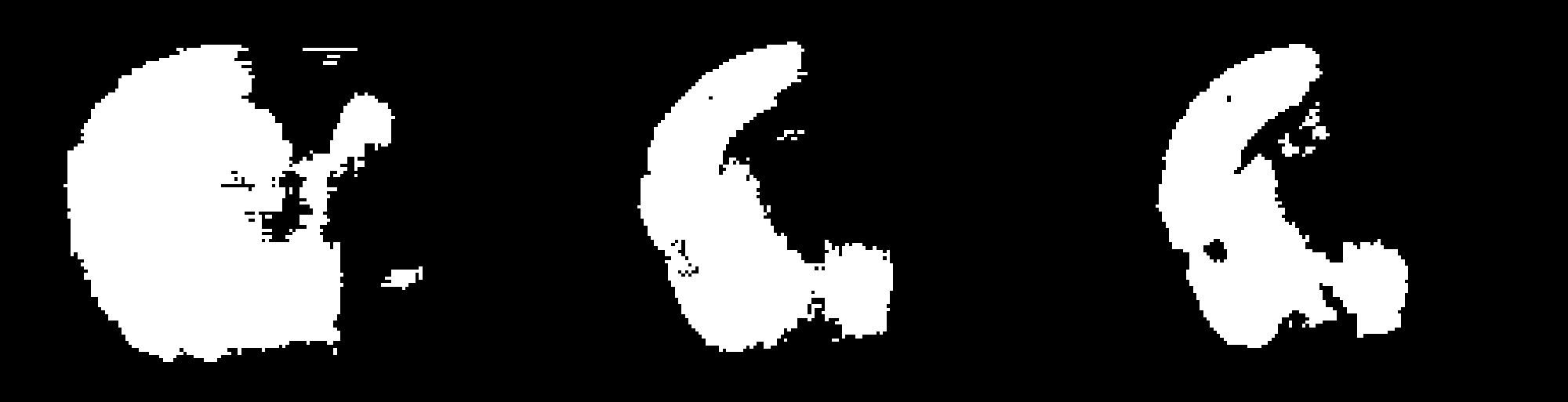}
\caption{Top: A slice of a T2-weighted rat brain image with lesion and its segmentation. Bottom: segmentations generated by 3D U-Net, VoxResNet and our architecture in the cross-validation test.} \label{predictions}
\end{figure}

The segmentations generated with RatLesNet have higher Dice coefficients than inter-operator variability. Furthermore, predictions did not only achieve greater Dice coefficients than the other 3D networks, but also the quality of the final segmentations was noticeably better in visual inspection: Figure \ref{predictions} depicts the predicted lesion in the same slice of the brain generated by the three different architectures, together with the original MR image and its ground truth. 3D U-Net's segmentation shows that the architecture did not learn to generalize properly since it classifies large areas of unequivocal healthy tissue as lesion. The borders of VoxResNet's prediction were comparatively crispier than the borders of the segmentation produced by our architecture. RatLesNet generated consistent boundaries such as the gap between the two affected areas at the bottom of the slice and the exclusion of the vein in the left side of the cortical region. The consistency in the boundaries also increased the size of holes within the segmented lesion and the size of clustered lesion voxels outside the affected area.

\textbf{Generalization capability}. 3D U-Net, VoxResNet and the proposed architecture were trained on the scans from 02NOV2016 study (including sham animals) and tested on the scans from the remaining 3 studies to compare their generalization capabilities. Table \ref{table2} summarizes the results. Our approach produced markedly larger Dice coefficients than VoxResNet and 3D U-Net in 03MAY2016 and 02OCT2017 studies. On the other hand, VoxResNet architecture produced the largest Dice coefficients in 03AUG2015 study. With 03MAY2016 and 02OCT2017 studies, the Dice coefficients of our network were markedly larger than the Dice coefficient between two independent manual segmentations (0.73) of 02NOV2016 study.  

\begin{table}
\caption{Average Dice scores and standard deviation. The model was trained on the 02NOV2016 study.}\label{table2}
\centering
\begin{tabular}{|l|c|c|c|}
\hline
Study & \hspace{0.01in} 3D U-Net \hspace{0.01in} & \hspace{0.01in} VoxResNet \hspace{0.01in} & \hspace{0.01in} RatLesNet \hspace{0.01in} \\
\hline
03AUG2015 (D35) & \hspace{0.01in} $0.64 \pm 0.27$ \hspace{0.01in} & \hspace{0.01in} $0.71 \pm 0.23$ \hspace{0.01in} & \hspace{0.01in} $0.68 \pm 0.26$ \hspace{0.01in} \\
03MAY2016 (24h) & \hspace{0.01in} $0.62 \pm 0.18$ \hspace{0.01in} & \hspace{0.01in} $0.77 \pm 0.12$ \hspace{0.01in} & \hspace{0.01in} $0.82 \pm 0.05$ \hspace{0.01in} \\
02OCT2017 (24h) & \hspace{0.01in} $0.60 \pm 0.18$ \hspace{0.01in} & \hspace{0.01in} $0.78 \pm 0.07$ \hspace{0.01in} & \hspace{0.01in} $0.84 \pm 0.04$ \hspace{0.01in} \\ \hline
Average & \hspace{0.01in} $0.62 \pm 0.21$ \hspace{0.01in} & \hspace{0.01in} $0.76 \pm 0.15$ \hspace{0.01in} & \hspace{0.01in} $0.79 \pm 0.15$ \hspace{0.01in} \\
\hline
\end{tabular}
\end{table}

\noindent \textbf{Computation time}: The network was implemented in Tensorflow \cite{tf} and it was run on a Ubuntu 16.04 with an Intel Xeon W-2125 CPU @ 4.00GHz processor, 64 GB of memory and an NVidia GeForce GTX 1080 Ti with 11 GB of memory. Training lasted for approximately 6 hours and inference time was about half a second per scan.

\section{Conclusion}
We have presented RatLesNet, the first FCN specifically designed for rat brain lesion segmentation. Our approach does not rely on corrections on the MR images or post-processing operations, and utilizes the entire 3D volume to explicitly use spatial information in three dimensions. In addition, the architecture of our RatLesNet is agnostic to the number of modalities and MR sequences presented in the input channels, so its generalization to multimodal data is technically straightforward. Five-fold cross-validation showed that our approach is more accurate than two other 3D FCNs designed for anatomical segmentation, and the boundaries consistency of the predicted segmentations is higher. Furthermore, the generalization capabilities of the architectures was assessed with data from three additional studies, and our approach provided with markedly larger Dice coefficients in two of the three studies. Its performance on the third study was slightly worse than the performance of VoxResNet. 

Future work will expand the dataset increasing the variability of the lesion's appearance by including brain scans at different time-points after the lesion is caused. Additional study is also needed to understand if the underperformance of the other two networks is caused by the comparatively large number of parameters as they are more prone to overfit the data.

\textbf{Acknowledgments.} J.M.V.'s work was funded from the European Union's Horizon 2020 Framework Programme (Marie Skłodowska Curie grant agreement \#740264 (GENOMMED)) and R.D.F.'s work was funded from Marie Skłodowska Curie grant agreement \#691110 (MICROBRADAM). We also acknowledge the Academy of Finland grants (\#275453 to A.S. and \#316258 to J.T.).

\bibliographystyle{splncs04}
\bibliography{manuscript}

\begin{thebibliography}{10}
\providecommand{\url}[1]{\texttt{#1}}
\providecommand{\urlprefix}{URL }
\providecommand{\doi}[1]{https://doi.org/#1}

\bibitem{tf}
Abadi, M., Barham, P., Chen, J., Chen, Z., Davis, A., Dean, J., Devin, M.,
  Ghemawat, S., Irving, G., Isard, M., et~al.: Tensorflow: A system for
  large-scale machine learning. In: 12th $\{$USENIX$\}$ Symposium on Operating
  Systems Design and Implementation ($\{$OSDI$\}$ 16). pp. 265--283 (2016)

\bibitem{stat}
Arnaud, A., Forbes, F., Coquery, N., Collomb, N., Lemasson, B., Barbier, E.L.:
  Fully automatic lesion localization and characterization: Application to
  brain tumors using multiparametric quantitative mri data. IEEE transactions
  on medical imaging  \textbf{37}(7),  1678--1689 (2018)

\bibitem{voxresnet}
Chen, H., Dou, Q., Yu, L., Qin, J., Heng, P.A.: Voxresnet: Deep voxelwise
  residual networks for brain segmentation from 3d mr images. NeuroImage
  \textbf{170},  446--455 (2018)

\bibitem{thresholding}
Choi, C.H., Yi, K.S., Lee, S.R., Lee, Y., Jeon, C.Y., Hwang, J., Lee, C., Choi,
  S.S., Lee, H.J., Cha, S.H.: A novel voxel-wise lesion segmentation technique
  on 3.0-t diffusion mri of hyperacute focal cerebral ischemia at 1 h after
  permanent mcao in rats. Journal of Cerebral Blood Flow \& Metabolism
  \textbf{38}(8),  1371--1383 (2018)

\bibitem{3dunet}
{\c{C}}i{\c{c}}ek, {\"O}., Abdulkadir, A., Lienkamp, S.S., Brox, T.,
  Ronneberger, O.: 3d u-net: learning dense volumetric segmentation from sparse
  annotation. In: International conference on medical image computing and
  computer-assisted intervention. pp. 424--432. Springer (2016)

\bibitem{shortreview}
De~Feo, R., Giove, F.: Towards an efficient segmentation of small rodents
  brain: a short critical review. Journal of neuroscience methods  (2019)

\bibitem{dice1945measures}
Dice, L.R.: Measures of the amount of ecologic association between species.
  Ecology  \textbf{26}(3),  297--302 (1945)

\bibitem{braintumor}
Havaei, M., Davy, A., Warde-Farley, D., Biard, A., Courville, A., Bengio, Y.,
  Pal, C., Jodoin, P.M., Larochelle, H.: Brain tumor segmentation with deep
  neural networks. Medical image analysis  \textbf{35},  18--31 (2017)

\bibitem{resnet}
He, K., Zhang, X., Ren, S., Sun, J.: Deep residual learning for image
  recognition. In: Proceedings of the IEEE conference on computer vision and
  pattern recognition. pp. 770--778 (2016)

\bibitem{densenet}
Huang, G., Liu, Z., Van Der~Maaten, L., Weinberger, K.Q.: Densely connected
  convolutional networks. In: Proceedings of the IEEE conference on computer
  vision and pattern recognition. pp. 4700--4708 (2017)

\bibitem{adam}
Kingma, D.P., Ba, J.: Adam: A method for stochastic optimization. arXiv
  preprint arXiv:1412.6980  (2014)

\bibitem{koizumimodel}
Koizumi, J., Yoshida, Y., Nakazawa, T., Ooneda, G.: Experimental studies of
  ischemic brain edema. 1. a new experimental model of cerebral embolism in
  rats in which recirculation can be introduced in the ischemic area. Jpn J
  stroke  \textbf{8}, ~1--8 (1986)

\bibitem{levelsets}
Mulder, I.A., Khmelinskii, A., Dzyubachyk, O., de~Jong, S., Rieff, N., Wermer,
  M.J., Hoehn, M., Lelieveldt, B.P., van~den Maagdenberg, A.M.: Automated
  ischemic lesion segmentation in mri mouse brain data after transient middle
  cerebral artery occlusion. Frontiers in neuroinformatics  \textbf{11}, ~3
  (2017)

\bibitem{bratspaper}
Myronenko, A.: 3d mri brain tumor segmentation using autoencoder
  regularization. In: International MICCAI Brainlesion Workshop. pp. 311--320.
  Springer (2018)

\bibitem{unpooling}
Noh, H., Hong, S., Han, B.: Learning deconvolution network for semantic
  segmentation. In: Proceedings of the IEEE international conference on
  computer vision. pp. 1520--1528 (2015)

\bibitem{unet}
Ronneberger, O., Fischer, P., Brox, T.: U-net: Convolutional networks for
  biomedical image segmentation. In: International Conference on Medical image
  computing and computer-assisted intervention. pp. 234--241. Springer (2015)

\bibitem{quicknat}
Roy, A.G., Conjeti, S., Navab, N., Wachinger, C., Initiative, A.D.N., et~al.:
  Quicknat: A fully convolutional network for quick and accurate segmentation
  of neuroanatomy. NeuroImage  \textbf{186},  713--727 (2019)

\bibitem{micebrain}
Roy, S., Knutsen, A., Korotcov, A., Bosomtwi, A., Dardzinski, B., Butman, J.A.,
  Pham, D.L.: A deep learning framework for brain extraction in humans and
  animals with traumatic brain injury. In: 2018 IEEE 15th International
  Symposium on Biomedical Imaging (ISBI 2018). pp. 687--691. IEEE (2018)

\end{thebibliography}

\end{document}